\documentclass[twocolumn,aps,prl,longbibliography,floatfix,10pt]{revtex4-2}

\usepackage{natbib}
\usepackage{times}
\usepackage{graphicx}
\usepackage{amsfonts}
\usepackage{amsmath,amsthm,amssymb,mathrsfs}
\usepackage{dsfont}
\usepackage{xcolor}
\usepackage{microtype}
\usepackage{cancel}
\usepackage{bm}
\usepackage{siunitx}
\usepackage[colorlinks=true, allcolors=black, citecolor=blue, linkcolor=blue, urlcolor=blue]{hyperref}
\usepackage[integrals]{wasysym}

\allowdisplaybreaks

\DeclareMathOperator{\Tr}{Tr}

\DeclareMathOperator{\sinc}{sinc}

\DeclareMathOperator{\tr}{tr}

\renewcommand{\rm}[1]{\mathrm{#1}}
\newcommand{\grad}{\bm{\nabla}}
\renewcommand{\v}[1]{\mathbf{#1}}
\renewcommand{\vr}{\mathbf{r}}
\newcommand{\vk}{\mathbf{k}}
\newcommand{\vq}{\mathbf{q}}
\newcommand{\vp}{\mathbf{p}}
\newcommand{\vA}{\mathbf{A}}
\newcommand{\A}{\tilde{\vA}}
\newcommand{\vS}{\mathbf{S}}
\newcommand{\eps}{\epsilon}
\newcommand{\vs}{\bm{\sigma}}

\newcommand{\on}{\omega_n}
\newcommand{\On}{\Omega_n}

\newcommand{\Area}{\mathcal{A}}
\newcommand{\Vol}{\mathcal{V}}

\newcommand{\vF}{v_\mathrm{F}}

\newcommand{\su}{\uparrow}

\newcommand{\G}{G}

\usepackage[capitalize]{cleveref}

\begin{document}

\title{Strong photon coupling to high-frequency antiferromagnetic magnons via topological surface states}

\author{Henrik T. Kaarbø}
\thanks{These two authors contributed equally}
\author{Henning G. Hugdal}
\thanks{These two authors contributed equally}
\author{Sol H. Jacobsen}
\affiliation{Center for Quantum Spintronics, Department of Physics, NTNU, Norwegian University of Science and Technology, NO-7491 Trondheim, Norway}

\begin{abstract}
We show strong coupling between antiferromagnetic magnons and microwave cavity photons at both high and externally controllable magnon frequencies. Using the fully quantum mechanical path-integral method, we study an antiferromagnetic insulator (AFM) interfaced with a topological insulator (TI), taking Bi$_2$Se$_3$--MnSe as a representative example. We show that the mutual coupling of the spin-polarized surface states of the TI to both the squeezed magnons and the circularly polarized cavity photons results in a Chern-Simons term that activates the stronger electric, rather than magnetic, dipole coupling. Moreover, a squeezing-mediated enhancement of the coupling is achieved due to the unequal interfacial exchange coupling to the AFM sublattices, resulting in a coupling strength up to several orders stronger than for direct magnon-photon coupling. While direct cavity-AFM coupling has so far been limited in its applicability due to weak or low frequency coupling, this result may advance the utilization of high-frequency cavity magnonics and enable its incorporation into quantum information technology.  
\end{abstract}

\maketitle

\begin{figure}
 \includegraphics[width = \columnwidth]{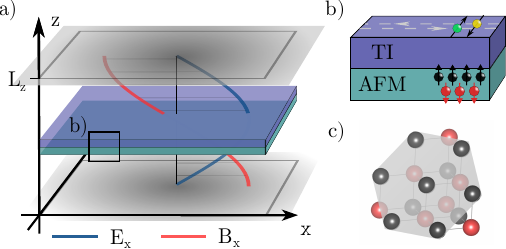}
\caption{\label{fig:system} (a) A bilayer comprising a thin AFM and TI is placed at the electric field maxima of an electromagnetic cavity. (b) The TI is interfaced to a spin uncompensated AFM surface, hosting spins from the sublattice pointing perpendicularly out of (black) or into (red) the interfacing plane. The spin-polarized TI surface states (green/yellow) couple to both AFM squeezed magnons and cavity photons. (c) Structure of the interfacing (111)-plane of the AFM bipartite sublattices.} 
\end{figure}

Strong coupling between magnons and microwave photons in cavities \cite{Huebl2013a,Tabuchi2014,Zhang2014,Bialek2021,Mergenthaler2017,Flower2019,Boventer2023} opens numerous possibilities for probing magnonic states \cite{Lachance-Quirion2017,Lachance-Quirion2020}, inducing unconventional spin states \cite{Chiocchetta2021}, and bridging spintronics and quantum information technology \cite{Tabuchi2015,Lachance-Quirion2019,ZareRameshti2022,Schlawin2022}. Information transfer and computational speeds are limited by the resonant frequency of the coupling, with device efficiency currently limited by the few GHz spin resonance frequencies of ferromagnetic magnons. The Zeeman coupling between isolated spins and cavity photons is in itself not strong, with coupling constant of the order of the Bohr magneton. Coupling to magnons, collective spin excitations, is enhanced compared to that of a single spin. Taken together with the confinement of electromagnetic waves in cavity systems, this gives rise to strong coupling in systems with macroscopic number of spins, such as ferromagnetic (FM) or antiferromagnetic (AFM) insulators. However, the reduced magnetic moment due to opposite spin alignment in AFMs significantly reduces their achievable magnetic dipolar coupling strength \cite{ZareRameshti2022}. Until now, strong coupling to AFMs has therefore only been shown for quasi-AFM phases with finite magnetic moments \cite{Bialek2021,Grishunin2018}, or at low frequencies \cite{Boventer2023}. Here, we use the path integral method to show that a TI-AFM bilayer can harness the potential of electric dipolar coupling (see \cref{fig:system}). By combining this with a large squeezing-mediated coupling enhancement from an uncompensated interface, we can achieve strong coupling at high frequencies.

The ratio between magnetic and electric dipolar coupling is proportional to the fine structure constant \cite{ZareRameshti2022}, meaning electric magnon-photon coupling has the potential to be stronger. However, the bound charges in FMs and AFMs do not couple resonantly to the electric cavity field to leading order. It is therefore generally overlooked as a coupling mechanism in cavities without driving, but has recently been shown to be relevant for photonic coupling to spin-flip excitations in metallic multi-band ferromagnets, where the magnons couple exclusively to either left- or right-handed circularly polarized cavity photons \cite{Hugdal2024}. 
Moreover, Lee \textit{et al} \cite{Lee2023a} recently proposed that the electric-field coupling can be boosted by an effective photon-magnon coupling achieved via a Chern-Simons (CS) term \cite{Altland2023,Nogueira2013,Rex2016,Rex2017} appearing when spin-polarized surface states of a topological insulator (TI) \cite{Hasan2010,Qi2011} mutually couple to both FM-magnons and cavity photons.  A CS term also appears in the effective theory of AFM magnons coupled to a TI, as long as the TI surface state dispersion mass gap is finite \cite{Rex2017,Nogueira2013}, achievable via magnetic doping \cite{Liu2009,Chen2010,Xu2012,Liu2012,He2017,Bhattacharyya2021} or an uncompensated AFM interface \cite{Eremeev2013,Luo2013,Eremeev2018,Bhattacharyya2021} [see \cref{fig:system}(b)]. Since easy-axis AFMs host magnons with two polarization axes \cite{Han2023},  they allow for selective coupling to circularly-polarized cavity photons, which we will show can be tuned and greatly boosted via interactions with a TI. 

Similarly to FM magnons, AFM magnons have unit spin. However, AFM magnon eigenstates do not correspond to spin flips in one of the sublattices, denoted sublattice magnons, but are rather superpositions of large numbers of sublattice magnons with total spin $\hbar$. The eigenstates are \emph{two-mode squeezed magnons} \cite{Kamra2019,Kamra2020}. Therefore, a particle with different interaction strengths to the two sublattice spins, e.g. via an uncompensated interface, can have an enhanced coupling to AFM magnons due to the dramatic increase in the effective magnon spin coupling to the particle -- a \textit{squeezing-mediated coupling enhancement} \cite{Kamra2019}, measurable via qubit excitation spectroscopy \cite{Romling2023,Romling2024}, for example. This has been shown to increase  superconducting- and exciton magnon-mediated pairing \cite{Erlandsen2019,Erlandsen2020,Erlandsen2020a,Thingstad2021,Johansen2019}, and may increase magnon-photon coupling in ferrimagnets \cite{Shim2020} and easy-axis FMs \cite{Lee2023b}. 

The cavity-bilayer system (\cref{fig:system}) is described by the action $S = S_\mathrm{cav}^0 + S_\mathrm{mag}^0 + S_\mathrm{TI}$, where interactions are included in $S_\rm{TI}$, and the bare cavity and magnon terms are ($\hbar = 1$) $     S_\mathrm{cav}^0 = -\sum_{q}\sum_{l=L,R} a_{ql}^\dagger (i\On - \omega_{\vq})a_{ql},$ and $S_\mathrm{mag}^0 = -\sum_{q} \sum_{l=L,R} \mu_{ql}^\dagger (i\On - \omega_{\vq l})\mu_{ql}.$ The bosonic fields $a_{ql}^{(\dagger)}$ describe left- (L) and right-handed (R) circularly polarized photons, and $\mu_{ql}^{(\dagger)}$ describe the two branches of squeezed magnons in a bipartite lattice \cite{Kamra2019}, which in easy-axis AFMs are L- and R-handed circularly polarized \cite{Han2023}. Since R (L) photons and magnons both have spin $+\hbar$ $(-\hbar)$ \cite{Beth1936,Kamra2017}, the parameter $l$ will denote polarization $(l=R/L)$ \emph{and} spin quantum number $(l=\pm 1)$. The three-vector $q = (\On, \vq)$, with momentum $\vq$ in the $xy$ plane and bosonic Matsubara frequency $\On = 2\pi n/\beta$, for $n\in \mathbb{Z}$ and $\beta = 1/k_\mathrm{B} T$, where $k_\mathrm{B}$ is the Boltzmann constant and $T$ the temperature \cite{Altland2023}. Assuming $L_{x,y}\gg L_z$, we focus on modes with $n_z = 1$, such that the cavity mode dispersion is $\omega_\vq = c\sqrt{\pi^2/L_z^2 + \vq^2}$, with speed of light $c$. For an AFM with easy-axis anisotropy $K$ experiencing a local static magnetic field $\bm{B}$ oriented along $\hat{z}$, the magnon branches are given by $\omega_{\vq l} = \sqrt{(1-\gamma_\vq^2)\omega_E^2 + \omega_K(2\omega_E + \omega_K)} + l \omega_H$ \cite{Johansen2018,Johansen2019a}. The characteristic frequencies of the exchange interaction, easy-axis anisotropy and magnetic field are $\omega_E =2SJ_{AB}N_\delta,\, \omega_K= 2SK$ and $\omega_H= g_J\mu_BB$ respectively, for spin $S$, intersublattice exchange parameter $J_{AB}$, the number of nearest neighbors $N_\delta$, Bohr magneton $\mu_B$, and Landé g-factor $g_J$. The structure factor
$\gamma_\vq = N_\delta^{-1}\sum_{\{\bm{\delta}\}} e^{i\vq\cdot \bm{\delta}}$, where $\{\bm{\delta}\}$ is the set of $N_\delta$ nearest neighbor lattice vectors \cite{Auerbach1994}. Notably, this means that the zero-momentum mode of the AFM is given by
\begin{align}
    \omega_{\vq=0l} = \sqrt{2\omega_E\omega_K + \omega_K^2 } + l\omega_H\label{eq:omega_magnon}.
\end{align}

We model the interlayer coupling between TI spins, $\bm{S}_{\mathrm{TI}}$, and AFM spins, $\bm{S}_{s}$, on sublattices $s = A,B$, using the Hamiltonian $H_{\mathrm{int}} = \sum_s J_s\bm{S}_{\mathrm{TI}}\cdot \bm{S}_{s}$, with exchange constants $J_s$. Thus, the TI action in imaginary time ($\tau$) and position ($\vr$) space, including interactions with the cavity and AFM is \cite{Rex2017,Hugdal2018b,Hugdal2020}
\begin{align}
    S_\mathrm{TI} = \int_0^\beta \!\! d\tau \!\int\!\! d\vr ~ \psi^\dagger \!\Big[\partial_\tau - (i\vF \grad + \tilde{\vA})\cdot\vs - m\sigma_z \Big] \!\psi, \label{eq:S_TI}
\end{align}
where $\psi = (\psi_\su~ \psi_\downarrow)^T$ with spin-up and -down fermion fields, $\vs = (\sigma_y, -\sigma_x)$ is the rotated 2D Pauli vector, $\vF$ is the TI's Fermi velocity, and the TI mass gap $m =J_A S_A^z + J_B S_B^z$ is due to the exchange interaction of the out-of-plane spin component of electrons in the neighboring AFM's $A$ and $B$ sublattice. An identity matrix is implicit in terms without explicit matrix structure. The interaction due to in-plane spin components has been combined with the in-plane cavity vector potential $\vA$ to define the effective vector potential
\begin{align}
    \tilde{\vA} ={}& \vF e\vA + \big(J_A \vS_A^\parallel + J_B \vS_B^\parallel\big)\times \hat{z}, \label{eq:A_eff}
\end{align}
with electron charge $e$, and unit vector $\hat{z}$.  $\vA$ fulfills the Coulomb gauge condition $\nabla \cdot \vA = 0$, and $\vS_s^\parallel$ and $S_s^z$ are the in-plane and out-of-plane components of the AFM spin distribution on sublattice $s$. Assuming that the sublattice spins are ordered in $\pm\hat{z}$ with magnitude $S$, we get $m = S(J_A - J_B)$ and can treat in-plane spin terms as fluctuations. The average cavity vector potential is also small for an undriven cavity, and contributions from $\tilde{\vA}$ can be treated as small compared to the remaining terms in the TI action, collected in the inverse Green's function
$\G^{-1} = -\partial_\tau + i\vF \grad\cdot\vs + m \sigma_z.$ The direct interaction between the cavity and AFM is neglected, since it is weak compared to the effective interaction induced by the TI surface states.

Performing the Gaussian integral over fermionic fields results in a term $-\Tr\ln(-\G^{-1} + \tilde{\vA}\cdot \vs)$ in the system action, which to leading order in $\G |\A|$ leads to an additional term in the combined photon and magnon action, $S_\rm{cav} + S_\rm{mag} = S_\rm{cav}^0 + S_\rm{mag}^0 + \delta S$ \cite{Altland2023,Rex2017}, with
\begin{align}
    \delta S
    ={}& \frac{1}{2} \sum_{kq} \tr \G_k \A_{q}\cdot \vs \G_{k-q} \A_{-q}\cdot \vs, \label{eq:delta_S}
\end{align}
where $\tr$ is the trace over fermionic spin states. 
We use the spin basis of the TI surface states and three-vector $k = (\on,\vk)$, with fermion Matsubara frequency $\on$ and momentum $\vk$. The TI Green's function becomes
\begin{align}
    \G_{k} ={}& \frac{i\on + \vF\vk\cdot \vs - m\sigma_z}{(i\on - \eps_\vk)(i\on + \eps_\vk)}, \label{eq:G_k}
\end{align}
with surface state dispersion relation $\eps_\vk = \sqrt{(\vF\vk)^2 + m^2}$. Performing the trace over spin states for the terms containing three Pauli matrices, using $\tr(\sigma_i \sigma_j \sigma_k) = 2i\eps_{ijk}$ with Levi-Civita symbol $\eps_{ijk}$, and the Matsubara and regular sums over $k$ in \cref{eq:delta_S} in the $\vF|\vq| \ll m$ limit yields a topological CS term \cite{Altland2023,Rex2016,Rex2017,Nogueira2013}
\begin{align}
    S_\rm{CS} ={}& \sum_q \frac{\Area\beta m}{8\pi \vF^2|m|} \On (\A_q \times \A_{-q})\cdot \hat{z},\label{eq:S_CS}
\end{align}
with TI-AFM interface area $\Area$. We neglect the additional Maxwell and anisotropy terms that appear when performing the matrix trace in \cref{eq:delta_S} \cite{Rex2017}, since they only renormalize the photon and magnon theories or lead to magnon-photon coupling terms weaker than the CS term. Moreover, the strength of the CS term does not renormalize or change under scale transformations \cite{Nogueira2015}. Therefore, the CS term introduces qualitatively new coupling channels with coupling constants that do not renormalize to zero. The effective vector potential [\cref{eq:A_eff}] in Matsubara and Fourier-space is
\begin{align}
    \A_q 
    ={}& \frac{i\vF e}{\sqrt{\eps_0 \omega_\vq \Vol\beta}
    } \sum_l \v{e}_l e^{il\phi_\vq}(a_{ql} + a_{-q\bar{l}}^\dagger) \nonumber\\*
    &+ i\sqrt{\frac{S}{N\beta}}\Big[ \v{e}_L (J_A m_{-qA}^\dagger + J_Bm_{qB}) \nonumber\\*
    &\qquad- \v{e}_R (J_Am_{qA} + J_B m_{-qB}^\dagger)\Big], \label{eq:A_eff_q}
\end{align}
where we have written $\vA$ in terms of the photon fields \cite{Kakazu1994,Janssonn2023} and performed the Holstein-Primakoff transformation for the spins \cite{Holstein1940}. Here, $\Vol= \Area L_z$ is the cavity volume, $m_{qA(B)}$ are the magnon fields for the $A(B)$ sublattice containing $N$ spins, and we have defined polarization vectors $\v{e}_{L/R} = (\hat{x} \pm i\hat{y})/\sqrt{2}$. The angle of $\vq$ relative to the $q_x  $ axis is $\phi_\vq$, and $\bar{l}$ denotes the opposite polarization of $l$. Since $\A$ contains both cavity and magnon fields, \cref{eq:S_CS} leads to corrections to both the cavity and spin theories, hereafter assumed to be included in $\omega_\vq$. Their cross terms, however, leave an effective coupling between the cavity and AFM spins, which we will elucidate.

The AFM action is diagonalized for the squeezed magnons by the Bogoliubov transformation \cite{Kamra2019,Johansen2019a}
\begin{align}
    \mu_{qR(L)} ={}& u_\vq m_{qA(B)} + v_\vq m_{-qB(A)}^\dagger \label{eq:Bogoliubov}
\end{align}
where $u_\vq(v_\vq) = \sqrt{\frac{\omega_E + \omega_K}{\omega_{\vq R} + \omega_{\vq L}} \pm \frac{1}{2}}$.
To find the magnon-photon coupling, we insert \cref{eq:A_eff_q,eq:Bogoliubov} into \cref{eq:S_CS}. Keeping only terms with both photon and magnon fields gives the interaction 
\begin{align}
    S_\rm{int} ={}& \sum_q i \On \kappa_\vq \bigg[\begin{pmatrix} \mu_{qR}^\dagger  & \mu_{-qL}\end{pmatrix} V_\vq \begin{pmatrix}
        a_{qR} \\
        a_{-qL}^\dagger
    \end{pmatrix} \nonumber\\*
    &\quad + \begin{pmatrix} a_{qR}^\dagger & 
        a_{-qL} \end{pmatrix} V_\vq^\dagger \begin{pmatrix}
        \mu_{qR} \\
        \mu_{-qL}^\dagger
    \end{pmatrix}\bigg]
    , \label{eq:S_int}
\end{align}
with dimensionless interaction matrix
\begin{align}
    V_\vq \!=\! [(u_\vq  \!-\! v_\vq)(1 \!+\! \sigma_x) \!+\! \Delta (u_\vq \!+\! v_\vq)(\sigma_z \!+\! i\sigma_y)]e^{i\phi_\vq}, \label{eq:V}
\end{align}
and 
$\kappa_\vq = -\frac{J\Area m e}{4\pi \vF |m|} \sqrt{\frac{S}{\eps_0 \omega_\vq N \Vol}}$, where $J = (J_A + J_B)/2$ and  $\Delta = (J_A - J_B)/(J_A + J_B).$

The interaction in \cref{eq:S_int,eq:V} displays two important properties and benefits of AFMs compared to FMs: (1) The coupling is proportional to the frequency $i\On \to \omega_{\vq l}$, which is orders of magnitude higher for AFMs; (2) If the TI's coupling to the two sublattices is unequal ($\Delta \neq 0$), the interaction strength can be increased dramatically. In the long wavelength limit, we have $u_\vq \pm v_\vq \sim (2\omega_E/\omega_K)^{\pm 1/4}$. For ratios $\omega_E/\omega_K \sim 10^{4}$ \cite{Kamra2019} it is therefore possible to increase the effective magnon-photon coupling by two orders of magnitude due to squeezing-enhancement when $\Delta \neq 0$, compared to equal sublattice coupling ($\Delta = 0$).

\begin{figure*}
\includegraphics[width=\textwidth, angle=0,clip]{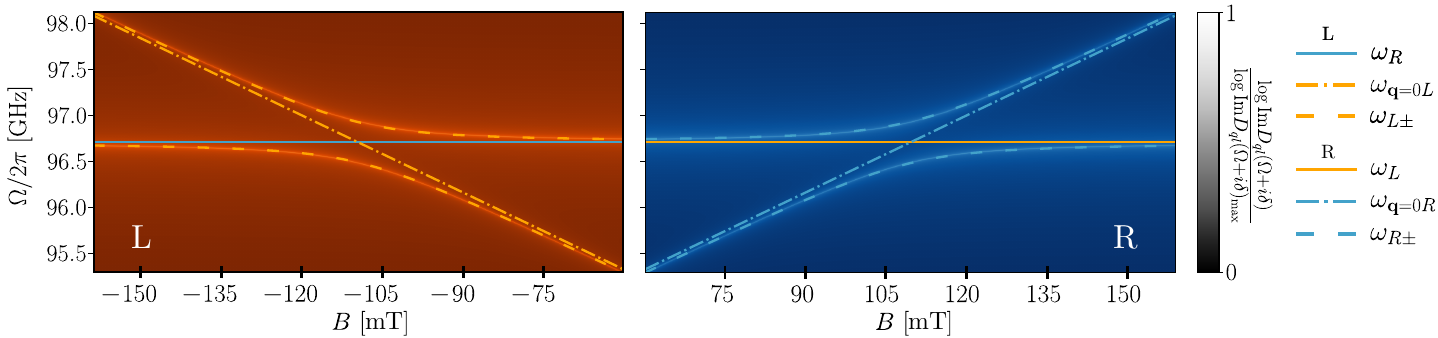}
    \caption{L- and R-photon spectral functions [\cref{eq:D}] versus applied magnetic field for $\omega_{\vq=0}/2\pi = \SI{96.7}{GHz}$, $\omega_E/\omega_K= 3\times 10^3$, corresponding to $\omega_{\vq=0,l}/2\pi = \SI{93.6}{\hertz} +l\omega_H/2\pi $, $\delta = \SI{e-15}{\radian\per\sec}$, and approximate solutions $\omega_{l\pm}$ [\cref{eq:omega_analytical}] for a Land\'e $g$-factor $g_J = 2$. The solid horizontal lines give the spectral function's peak for the \textit{opposite} polarization $\omega_{R(L)}$ in the given frequency range, which does not couple resonantly to the diagonally crossing dot-dashed magnon frequency $\omega_{\vq=0 L (R)}$ [\cref{eq:omega_magnon}], and therefore corresponds to the uncoupled photon resonances $\omega_{\vq=0}$. A gap of $\SI{0.5}{\giga\hertz}$ at resonance for the given chirality demonstrates strong coupling. System parameters are presented in the main text.}
    \label{fig:resonances}
\end{figure*}

The strong coupling due to these two mechanisms is illustrated by the resulting photon spectra of the system shown in \cref{fig:resonances}. The magnon-polariton resonances can be found combining \cref{eq:S_int} with the bare photon and magnon actions and setting the determinant of the resulting $4\times4$ matrix to zero and solving for $i\On$. However, since we are interested in both the resonances and spectral weights of the photons, relevant for e.g. input-output experiments, we integrate out the magnon modes \cite{Altland2023}, resulting in the effective cavity theory, $S_\rm{cav} = -\sum_q \begin{pmatrix}
        a_{qR}^\dagger & a_{-qL}
    \end{pmatrix}D^{-1}_q\begin{pmatrix}
        a_{qR} \\ a_{-qL}^\dagger
    \end{pmatrix}\!.$
Here, the inverse Green's function $D_q^{-1} = i\On\sigma_z - \omega_\vq - (1+\sigma_x)\Pi_q$
whose self-energy due to magnon interactions is 
\begin{align}
    \Pi_q ={}& -\On^2 \bigg\{\frac{g_{\vq R}^2}{i\On - \omega_{\vq R}} - \frac{g_{\vq L}^2}{i\On + \omega_{\vq L}}\bigg\}, \label{eq:Pi}
\end{align}
where
\begin{align}
    g_{\vq l} ={}& |\kappa_\vq||u_\vq-v_\vq +l \Delta (u_\vq + v_\vq)|. \label{eq:g}
\end{align}
Inverting $D_q^{-1}$, we find the R- and L-photon Green's functions \cite{Hugdal2024}
\begin{align}
    D_{q l} = {}& \frac{i \On + \omega_\vq + \Pi_{l q}}{(i\On)^2 - \omega_\vq^2 - 2\omega_\vq \Pi_{l q}}, \label{eq:D}
\end{align}
where the self-energy is evaluated at momenta $lq = \pm q$ for R/L-modes.

Photon dispersions in the interacting system are found by analytically continuing to real frequencies, $i\On \to \Omega + i\delta$, where $\delta = 0^+$, and finding the zeros of the denominator in \cref{eq:D}. The resulting root equation is fourth order in $\Omega$, but can be reduced to a cubic equation by disregarding negative energy solutions \footnote{Negative energy solutions correspond to the time-reversed process, which exchanges the role of the two sublattice magnons, as well as the photon polarization. This can be seen by exchanging $\Omega \to -\Omega$ in \cref{eq:Pi} and seeing how this is equivalent to exchanging the left and right polarization coupling constants}, i.e., letting $\Omega + \omega_\vq \to 2\omega_\vq$. Moreover, only one of the two terms in \cref{eq:Pi} is resonant and hence contributes significantly depending on the sign of $q$ in \cref{eq:D}, as one of the denominators vanishes. Neglecting the far off-resonant term, we get two solutions corresponding to the magnon-polariton modes for each polarization,
\begin{align}
    \omega_{l\pm} ={}& \frac{\omega_\vq + \omega_{\vq l}}{2(1-g_{\vq l}^2)} \pm \frac{\sqrt{(\omega_\vq - \omega_{\vq l})^2 + 4\omega_\vq\omega_{\vq l}g_{\vq l}^2}}{2(1-g_{\vq l}^2)}.\label{eq:omega_analytical}
\end{align}
For $|\omega_\vq - \omega_{\vq l}| \gg 2g_{\vq l}\sqrt{\omega_\vq\omega_{\vq l}}$ and $\omega_\vq > \omega_{\vq l}$ ($\omega_\vq < \omega_{\vq l}$), the solution with upper (lower) sign approaches $\omega_\vq$, while the lower (upper) sign solution approaches $\omega_{\vq l}$. For $\omega_\vq = \omega_{\vq l}$, the frequency splitting between the resonances is $2\omega_{\vq} g_{\vq l}\equiv 2\tilde{g}_{\vq l}$ %
meaning that the effective coupling strength between magnons and photons with polarization $l$ is approximately 
\begin{align}
    \tilde{g}_{\vq l} \approx{}&  \frac{em(u_\vq + v_\vq)}{8\pi \vF}\sqrt{\frac{\omega_\vq \Area_\rm{uc}}{\eps_0 S L_z}} 
    \approx \frac{em}{4\pi \vF}\sqrt{\frac{\omega_E \Area_\rm{uc}}{2\eps_0 S L_z}}, \label{eq:g_eff} 
\end{align}
for $\omega_K, \omega_H \ll \omega_E$, with $\Area_\rm{uc} = \Area/N$ the AFM unit cell area.
In the case $\Delta = 1$, substituting $( u_\vq + v_\vq) \to 2u_\vq\,(2v_\vq)$ in \cref{eq:g_eff} gives $g_{\vq R} (g_{\vq L})$. 
In the limit $\vq \to 0$ and $\omega_K, \omega_H \ll \omega_E$, this difference is negligible.
Since the cavity-TI coupling is proportional to $\vF$, while the effective cavity-magnon coupling depends on the density of fermion states in the TI, proportional to $\vF^{-2}$, $\tilde{g}_{\vq l}$ scales with $\vF^{-1}$.

The above example considered the AFM and TI area $\Area$ to be equal to the in-plane cavity area $\Area_\mathrm{cav}$. If instead $\Area\ll\Area_\mathrm{cav}$, the photons couple only to the uniform magnon modes. The interaction strength then gets an additional factor $\sqrt{\Area/\Area_\mathrm{cav}} \propto \sqrt{N}$, see the Supplemental Material \cite{supp}, as often reported for magnon-photon coupling \cite{Tabuchi2014,Zhang2014,Johansen2018,Flower2019,Lee2023a}. The last approximation elucidates the advantage of using uncompensated AFMs versus regular ferromagnets. At resonance, the coupling strength effectively depends on the characteristic exchange frequency $\omega_E$, which is typically an order of magnitude larger than the uniform mode frequencies for AFMs, and several orders of magnitude larger than its FM counterpart. This coupling is most readily compared to the magnon-photon coupling $g_{\vq\mathrm{Z}}$ based on the Zeeman interaction in antiferromagnets, given by $g_{\vq\mathrm{Z}} = \frac{|\gamma|}{c}\sqrt{\frac{\omega_{\vq}S}{\epsilon_0 \Area_\rm{uc} L_z}}(u_\vq - v_\vq)$ \cite{Johansen2018}, with gyromagnetic ratio $\gamma$.
The relative enhancement is thus 
\begin{equation}
    \Gamma_{\vq} = \frac{\tilde{g}_{\vq}}{g_{\vq\mathrm{Z}}} =\frac{emc\Area_\rm{uc}}{8\pi\vF S|\gamma|}\frac{(u_\vq + v_\vq)}{(u_\vq-v_\vq)} \approx \frac{emc\Area_\rm{uc}}{8\pi \vF S |\gamma| }\sqrt{\frac{2\omega_E}{\omega_K}},
\end{equation}
where the last approximation is valid when $\omega_E \gg \omega_K$.

The simplified expression for the coupling strength [\cref{eq:g_eff}] provides insight into the dominant mechanisms that control the coupling strength, and is valid close to resonance, when $\Delta$ is finite, $g_{\vq l} \ll 1$ and $\omega_E \gg \omega_K$. Applying the opposite limit, $\omega_E \ll \omega_K,$ using \cref{eq:g} yields that the squeezing-mediated enhancement due to the Bogoliubov coefficients $u_\vq$ and $v_\vq$ approaches $1+l\Delta$. For $l = +1$, the factor is monotonic in $\omega_K$, meaning that uncompensated squeezing \textit{always} enhances the coupling strength. As $g_{\vq l}$ approaches $1$, the system enters the ultra-strong or deep-strong regimes \cite{ZareRameshti2022}, whereby the Green's function \eqref{eq:D} is needed to accurately describe the spectra. 

To estimate the coupling strength, we consider the experimentally well-studied TI-AFM bilayer Bi$_2$Se$_3$--MnSe \cite{Luo2013,Eremeev2013,Matetskiy2015,Pournaghavi2021}. An alternative could be Bi$_2$Se$_3$--CrI$_3$ \cite{Hou2019a}, where bilayer CrI$_3$ is shown to have antiferromagnetically coupled FM layers \cite{Huang2017,Sivadas2018,Soenen2023}. For the TI we use $\vF = \SI{3e5}{\meter/\second}$ \cite{Luo2013,Zhang2009} and $m =\SI{45}{\milli\electronvolt}$ \cite{Matetskiy2015,Eremeev2013}, and for the AFM we set $S=2.5$ \cite{OHara2018}, $\Area_\rm{uc} = \SI{0.13}{\nano\meter\squared}$
\cite{Matetskiy2015}, $\omega_E/\omega_K \sim \numrange{e3}{e4}$ \cite{Sattar2022,Kamra2019} and $N_\delta = 3$, corresponding to the number of nearest neighbors in a crystal consisting of a single (111)-plane of a bipartite AFM.
The resulting relative enhancement of the magnon-photon interaction is $\Gamma_{\vq = 0}\sim\numrange{50}{170}$, showing that the coupling strength is comfortably enhanced by up to two orders. In an anisotropy-dominated AFM, where $\omega_K\gg \omega_E$, the enhancement would instead be of the order $\Gamma_\vq\sim 1.2\cdot(1+l\Delta)$.

\cref{fig:resonances} shows avoided crossings in the L/R-polarized photon dispersion curves as the characteristic signature of strong coupling to the different magnon branches with varying static external magnetic field. As an example, we set $L_z = \SI{1.55}{\milli\meter}$ for the $n_z = 1$ cavity mode. The external field can then be tuned to the uniform mode of an AFM with exchange parameter $|J_{AB}| = \SI{1}{\milli\eV}$. The specific value of $|J_{AB}|$ does not affect the coupling enhancement when $\omega_E/\omega_K$
remains constant. The approximate gap of $\SI{0.5}{\giga\hertz}$ at resonance seen in \cref{fig:resonances} corresponds to an effective coupling strength $\tilde{g}/2\pi \sim \SI{0.25}{\giga\hertz}$. The cooperativity $C = \tilde{g}^2/\kappa\gamma$ \cite{ZareRameshti2022} is above unity even for pessimistic values for the cavity and magnon relaxation rates, $\kappa, \gamma \sim \SI{100}{\mega\hertz}$, and $C > 1000$ is possible for high-quality cavities and AFMs \cite{Mergenthaler2017,Boventer2023}. This prediction is therefore in the strong coupling regime.

The ultra-strong and deep-strong regimes correspond to $\tilde{g} \lesssim \Omega$ and $\tilde{g} \gtrsim \Omega$, 
respectively, where $\Omega$ is the mode frequency \cite{ZareRameshti2022}. From \cref{fig:resonances} we find $\tilde{g}/\Omega \approx \num{2.6e-3}$. \cref{eq:g} indicates that the ultra-strong limit, two orders stonger than the current regime, may be approached by increasing the ratio $J/\vF$, i.e., using an AFM with high exchange coupling $J$ and decreasing the TI Fermi velocity, for instance by impurity doping \cite{Wolos2012} or an applied uniform electric field \cite{Diaz-Fernandez2017}.

The key dependency for resonant coupling is the magnon-cavity frequency matching condition. As shown in \cref{fig:resonances}, the split dependence of the uniform magnon modes' frequencies on the static external magnetic field enables specific chirality selection. A modulation of the external field could then switch transmission of an unpolarized photon signal between magnon branches, acting as a signal splitter, or frequency-dependent linear-to-circular polarization converter \cite{Wang2017a}, and enable non-invasive cavity engineering \cite{Hubener2021,Schlawin2022,Hugdal2024}. A given magnon branch's frequency can therefore be tuned by an external field. The available frequencies are limited above by the AFM's spin-flop transition when $\omega_H = \omega_{\vq l}$, when out-of-plane ordering is lost \cite{ZareRameshti2022}; the Hamiltonian for a TI has been experimentally shown to hold for fields of at least $11$ T \cite{Cheng2010,Hanaguri2010}. As such, an external field may induce resonant coupling towards the \SI{}{\tera\hertz}-regime, limited by $2\omega_{\vq = 0}$. Furthermore, viewing linearly polarized light as a superposition of circularly polarized light, the chirality-selective coupling could be detected as a frequency dependent Kerr-rotation of the incoming light as the opposing polarization is unaffected by the system.

The analytic expressions [\cref{eq:omega_analytical}] and numerical results [using \cref{eq:D}] shown in \cref{fig:resonances} are in excellent agreement. The numerical results do not explicitly include the correction to the magnon dispersion due to interaction with the TI, e.g. from the magnon-magnon terms in \cref{eq:S_CS}, which would in practice result in a small, degeneracy lifting renormalization of the magnonic frequency \cite{Rex2017}. When $J_A\gg J_B$, this renormalization is $\Omega_{nl} = \Omega_n\left[1-lu_\vq^2(v_\vq^2)\zeta\right]$ for the dimensionless magnon-magnon coupling constant $\zeta = \frac{SJ_A^2\Area_\rm{u.c.}}{8\pi\vF^2}$\footnote{The renormalization breaks the boson-characteristic $2\pi n/\beta$-periodicity of the magnonic Matsubara frequencies $\Omega_n$. This suggests that the magnons take on an anyonic character in the two-dimensional limit}.
For the given system parameters, $\zeta \sim 10^{-4}$, corresponding to energy on the order of \SI{5e-2}{\micro\eV} at \SI{100}{\giga\hertz}. The effective magnon-magnon interaction therefore results in a negligible change in the AFM magnon squeezing.

We emphasize that the strong coupling enhancement predicted here is an interface effect, which cannot be further enhanced by adding magnetic layers in the bulk (increasing $N$ beyond the limit set by $\Area \lesssim \Area_\mathrm{cav}$) as is often employed for maximizing magnetic dipole coupling. However, as $g_{\vq Z} \sim \sqrt{N/\Vol} \sim 1/\sqrt{\Area_\rm{uc} L_z}$, the coupling increase is not due to the increased numbers of spin directly, but rather the increase in geometrical overlap factor when increasing the magnetic volume (see Supplemental Material \cite{supp} for detailed calculation). The geometrical overlap factor would need to be four orders of magnitude greater for the standard Zeeman effect to rival the described interface interaction. Moreover, the overlap factor can be increased by cavity engineering, relevant for both coupling mechanisms, in addition to increasing the magnet volume \cite{Flower2019}. Unlike bulk Zeeman-based coupling, which requires that the photon and magnon wavelengths coincide across the bulk to generate coherent magnons \cite{Salikhov2023}, the bilayer acts as a strong photomagnonic spin pump at the AFM/TI interface. This is advantageous, as the generated magnon only needs to be in-phase with the photon at the interface, before propagating into a neighboring magnetic material \cite{Salikhov2023}. The coupling mechanism introduced here may therefore enable photonic control of coherent excitations in ultrafast devices, especially in cases where the AFM layer is scaled down for on-chip applications.

Strong coupling allows mutual exchange and manipulation of magnonic and photonic information. Its potential could be further developed by exploiting the external control for signal splitting, e.g. via a quasi-static oscillating field, and the possibility of controlling other time-reversal symmetry-breaking systems which couple to only one photon mode. In addition, one can investigate the consequences for strong coupling to qubits \cite{Romling2024}, or superconducting reservoirs, which also couple via the electric field \cite{Janssonn2020,Janssonn2023}. The same coupling mechanism is expected to apply to ferrimagnets. Materials with strong spin-orbit coupling, and coupling via other spin Hall effects \cite{Sinova2015,Kamra2023}, may provide interesting alternatives to the TI. However, since the appearance of a CS term is linked to nontrivial band topology \cite{Qi2008,Nogueira2015}, non-topological materials with strong spin-orbit coupling may display qualitatively different and less robust magnon-photon couplings. Furthermore, the higher Néel temperatures of metallic AFMs could also enable strong coupling in bilayer setups. 
Moreover, since charge-transport can reposition the Néel vector between the gapped and gapless phases in topological metallic AFMs \cite{Siddiqui2020}, a similar setup could provide an externally tunable way to manipulate the AFM-ordering, and provide a potential mechanism for dual read-write protocols.

We have shown strong, tunable, high-frequency coupling between AFM magnons and microwave cavity photons. This is mediated by a %
Chern-Simons term appearing due to the mutual coupling of spin-polarized TI surface states to both the squeezed magnons and cavity photons. This activates the strong electric dipole coupling, which is up to several orders stronger than direct magnon-photon coupling. Experimental confirmation of this prediction would be a significant step towards the ambition of incorporating high-frequency magnonics with photonic quantum information technologies.

\appendix
\section*{Appendix: Detailed derivation of magnon-photon coupling and its limiting cases}\label{Sec:Appendix}

The two-dimensional, real-space TI action [Eq. (2) in main text] is Fourier transformed using the relations
\begin{subequations}
\begin{align}
    \psi ={}& \frac{1}{\sqrt{\Area_\mathrm{TI}\beta}} \sum_k \psi_k e^{-ikr},\\
    \psi^\dagger ={}& \frac{1}{\sqrt{\Area_\mathrm{TI}\beta}} \sum_k \psi_k^\dagger e^{ikr}.
\end{align}
\end{subequations}
This results in the TI action
\begin{align}
    S_\mathrm{TI} = \sum_{kk'}\psi_k^\dagger [G_k^{-1}\delta_{kk'} - \tilde{\vA}_{k-k'}\cdot \vs]\psi_{k'},\label{eq:STI_app}
\end{align}
where the surface state propagator in Matsubara frequency and momentum space is \cite{Hugdal2020} 
\begin{align}
    G_k^{-1} ={}& -i\on + \vF\vk\cdot\vs - \alpha m \sigma_z,
\end{align}
and $\alpha = \Area_\textrm{AFM-TI}/\Area_\textrm{TI} \leq 1$ is the ratio of the AFM-TI interface area to the total TI in-plane area. 

The effective gauge field in \cref{eq:STI_app} is given by
\begin{align}
    \tilde{\vA}_{k-k'} ={}& \frac{1}{\Area_\mathrm{TI}\beta }\int_0^\beta d\tau \int d\vr ~\tilde{\vA} e^{i(k-k')r} \nonumber\\*
    ={}& \sum_q \bigg\{\frac{i\vF e}{\sqrt{\eps_0 \omega_\vq \Vol \beta}} \sum_l \v{e}_l e^{il\phi_\vq}(a_{ql} + a_{-q\bar{l}}^\dagger) D_{k-k',q}^\mathrm{cav}\nonumber\\*
    &+ i\sqrt{\frac{S}{N\beta}} [\v{e}_L(J_A m_{-qA}^\dagger + J_B m_{qB}) \nonumber\\*
    &\quad- \v{e}_R (J_Am_{qA} + J_Bm_{-qB}^\dagger)] D_{k-k',q}^\mathrm{AFM}\bigg\} \nonumber\\*
    \equiv{}& \sum_q \big[\vA_q^\mathrm{cav} D_{k-k',q}^\mathrm{cav} + \vA_q^\mathrm{AFM}D_{k-k',q}^\mathrm{AFM}\big]. \label{eq:A_tilde_unequal}
\end{align}
Here we have defined the quantity \cite{Janssonn2023}
\begin{align}
    D_{kq}^I ={}& \frac{\delta_{\on\On}}{\Area_\mathrm{TI}} \int\limits_{\Area_\cap} d\vr ~ e^{i(\vk-\vq)\cdot\vr} \nonumber\\*
    ={}& \frac{\Area_\cap}{\Area_\mathrm{TI}} \delta_{\on\On}  \prod_{i=x,y} \sinc\left(\frac{k_i - q_i}{2}l_i^I\right) \nonumber\\*
    ={}& \begin{cases}
        \delta_{\on\On} \!\! \prod\limits_{i=x,y} \!\!\sinc\!\!\big[\pi\!\big(n_i - j_i\frac{l_i^\mathrm{TI}}{L_i}\big)\!\big]\!, & I = \text{cav},\\
        \alpha\delta_{\on\On}\!\! \prod\limits_{i=x,y}\!\! \sinc\!\!\big[\pi \!\big(\!\frac{ n_i l_i^\cap}{l_i^\mathrm{TI}} - \frac{ j_i l_i^\cap}{l_i^\mathrm{AFM}}\big)\!\big]\!, & I = \text{AFM},
    \end{cases}\label{eq:D_App}
\end{align}
where $n_i, j_i \in \mathbb{Z}$ denote the momentum indices. We have assumed that the center of the AFM and TI is placed in the at the origin, and that we have an interface area $\Area_I = l_x^I l_y^I$. For $I = \mathrm{cav}$, we have $\Area_\cap = \Area_\mathrm{TI}$, since the TI has to be contained within the cavity, i.e. $L_i \geq l_i^\mathrm{TI}$. This criterion could potentially be relaxed for analogous on-chip/resonator systems. For $I = \mathrm{AFM}$, the area is determined by the physical interface area, which for simplicity is assumed to be square with lengths $l_i^\cap = \min(l_i^\mathrm{TI}, l_i^\mathrm{AFM})$.

$D_{k,q}^I$ reduces to a Kronecker delta in momentum only when the subsystem dimensions are (approximately) identical. Large differences in subsystem length along direction $i$ leads to coupling between all momentum modes along $i$ in the larger subsystem, but only to the uniform mode of the smaller subsystem.

The partition function for the total system is \cite{Altland2023,Janssonn2023}
\begin{align}
    Z = \int \!\! D[\psi^\dagger,\psi]  \int \!\! D[m^\dagger, m] \int \!\! D[a^\dagger, a] ~ e^{-S_\mathrm{TI} - S_\mathrm{AFM}^0 - S_\mathrm{cav}^0},
\end{align}
where, for instance, $\int D[\psi, \psi^\dagger]$ denotes the path integral over every TI surface state mode.
Performing the Gaussian path integral over the TI surface states described by \cref{eq:STI_app} \cite{Rex2017,Altland2023}, the leading order correction is
\begin{align}
    \delta S ={}& \frac{1}{2}\sum_{kk'}\tr G_k \tilde{\vA}_{k-k'}\cdot \vs G_{k'}\tilde{\vA}_{k'-k}\cdot \vs \nonumber\\*
    ={}& \frac{1}{2}\sum_{kpqq'}\tr G_k \big[\vA_q^\mathrm{cav} D_{pq}^\mathrm{cav} + \vA_q^\mathrm{AFM}D_{pq}^\mathrm{AFM}\big]\cdot \vs \nonumber\\*
    &\times G_{k-p} \big[\vA_{-q'}^\mathrm{cav} D_{pq'}^\mathrm{cav} + \vA_{-q'}^\mathrm{AFM}D_{pq'}^\mathrm{AFM}\big]\cdot \vs,
\end{align}
where we have introduced $p = k-k'$, and $G_k$ is given in Eq. (5) in the main text, with $m \to \alpha m$. Performing the matrix trace and the sum over $k$, we arrive at the CS term [c.f. Eq. (6) in main text]
\begin{align}
    S_\mathrm{CS} = {}& \sum_{pqq'} \frac{\Area_\mathrm{TI} \beta m}{8\pi\vF^2|m|} \On \Big[\big(\vA_q^\mathrm{cav} D_{pq}^\mathrm{cav} + \vA_q^\mathrm{AFM}D_{pq}^\mathrm{AFM}\big)\nonumber\\*
    &\times \big(\vA_{-q'}^\mathrm{cav} D_{pq'}^\mathrm{cav} + \vA_{-q'}^\mathrm{AFM}D_{pq'}^\mathrm{AFM}\big)\Big]\cdot \hat{z}. \label{eq:S_CSapp}
\end{align}
Focusing on the magnon-photon interaction as in the main text, we keep only terms involving both the AFM and cavity, resulting in the interaction [c.f. Eq. (9) in main text]
\begin{align}
    S_\mathrm{int} = {}& \sum_{qq'} i\On \kappa_{\vq} \bigg[\begin{pmatrix}
        \mu_{q'R}^\dagger & \mu_{-q'L}
    \end{pmatrix} V_{q'q} \begin{pmatrix}
        a_{qR} \\ a_{-qL}^\dagger
    \end{pmatrix} \nonumber\\*
    &+ \begin{pmatrix}
        a_{qR}^\dagger & a_{-qL}
    \end{pmatrix}V_{q'q}^\dagger\begin{pmatrix}
        \mu_{q'R} \\ \mu_{-q'L}^\dagger
    \end{pmatrix} \bigg],
\end{align}
where
\begin{align}
    \kappa_\vq ={}& - \frac{J \Area_\mathrm{TI} me}{4\pi\vF |m|} \sqrt{\frac{S}{\eps_0 \omega_\vq N \Vol}},
\end{align}
and 
\begin{align}
    V_{qq'} ={}& \sum_p D_{pq}^\mathrm{AFM}D_{pq'}^\mathrm{cav}e^{i\phi_{\vq'}}\nonumber\\*
    &\times[(u_{\vq} - v_{\vq})(1 + \sigma_x) + \Delta (u_\vq + v_\vq)(\sigma_z + i \sigma_y)]. \label{eq:Vqq}
\end{align}

To simplify the analysis, we use the rotating wave approximation. In the main text, we integrate out the magnons and find the interaction strength from the magnon-polariton dispersion at resonance. The rotating wave approximation can also be used to give an accurate estimate for sufficiently weak coupling, effectively neglecting the off-diagonal terms in \cref{eq:Vqq}. We then arrive at a coupling strength between photons and magnons with spin $l$ and momentum $\vq$ and $\vq'$, respectively, interacting at resonance frequency $\omega$:
\begin{align}
    \tilde{g}_{\vq\vq'l} = \sum_{\v{p}} \omega |\kappa_\vq| D_{\v{p}\vq}^\mathrm{cav} D_{\v{p}\vq'}^\mathrm{AFM} |u_{\vq'} - v_{\vq'} + l\Delta(u_{\vq'} + v_{\vq'})|,
\end{align}
where $D_{\vp\vq}^I$ denotes the spatial part of $D_{pq}^I$ in \cref{eq:D_App}.

We now consider three limiting cases: (i)  equal subsystem areas ($\Area_\mathrm{cav} = \Area_\mathrm{TI} = \Area_\mathrm{AFM}$), (ii) small TI and small AFM ($\Area_\mathrm{cav} \gg \Area_\mathrm{TI} = \Area_\mathrm{AFM}$), and (iii) small AFM but large TI ($\Area_\mathrm{cav} = \Area_\mathrm{TI} \gg \Area_\mathrm{AFM}$).

\paragraph{(i) Equal subsystem areas.} When $\Area_\mathrm{cav} = \Area_\mathrm{TI} = \Area_\mathrm{AFM}$, the quantities $D_{\vp\vq}$ [\cref{eq:D_App}] entering \cref{eq:S_CSapp} becomes $D_{\vp\vq}^I = \delta_{\vp\vq}$ for all $I$, such that 
\begin{align}
    \tilde{g}_{\vq\vq'l}^{(i)} = \tilde{g}_{\vq l} \delta_{\vq\vq'} =  \omega |\kappa_\vq| |u_\vq - v_\vq + l \Delta(u_\vq + v_\vq)| \delta_{\vq\vq'}.
\end{align}
in agreement with the results in the main text at resonance $\omega = \omega_\vq$.

\paragraph{(ii) Small TI and AFM.} For identical TI and AFM areas, both much smaller than the cavity in-plane area, $\Area_\mathrm{cav} \gg \Area_\mathrm{TI} = \Area_\mathrm{AFM}$, we have $D_{\vp\vq}^\mathrm{cav} \approx \delta_{0\vp}$ and $D_{\vp\vq}^\mathrm{AFM} = \delta_{\vp\vq}$. This leads to the coupling
\begin{align}
    \tilde{g}_{\vq\vq'l}^{(ii)} = \omega |\kappa_\vq| |u_0 - v_0 + l\Delta(u_0 + v_0)| \delta_{\vq'0},
\end{align}
that is, the photon modes only couple to the uniform ($\vq' = 0$) magnon modes.

\paragraph{(iii) Small AFM and large TI.} When the AFM is much smaller than both the cavity in-plane area and the TI, $\Area_\mathrm{cav} = \Area_\mathrm{TI} \gg \Area_\mathrm{AFM}$, we have $D_{\vp\vq}^\mathrm{cav} = \delta_{\vp\vq}$ and $D_{\vp\vq}^\mathrm{AFM} \approx \frac{\Area_\mathrm{AFM}}{\Area_\mathrm{TI}}\delta_{\vq 0}$, resulting in 
\begin{align}
    \tilde{g}_{\vq\vq'l}^{(iii)} = \alpha\omega |\kappa_\vq| |u_0 - v_0 + l\Delta(u_0 + v_0)|\delta_{\vq'0}.
\end{align}
This coupling is seemingly smaller than $\tilde{g}_{\vq\vq'l}^{(ii)}$ by a factor $\alpha = \Area_\mathrm{AFM}/\Area_\mathrm{TI}$. However, it is important to note that $\kappa_{\vq} \propto \Area_\mathrm{TI}$, which in case (ii) is much smaller than $\Area_\mathrm{cav}$. Hence, since $\tilde{g}_{\vq\vq'l}^{(ii)} \propto \Area_\mathrm{TI}$, and $\tilde{g}_{\vq\vq'l}^{(iii)} \propto \Area_\mathrm{AFM}$, we have $\tilde{g}_{\vq\vq'l}^{(ii)} = \tilde{g}_{\vq\vq'l}^{(iii)} $ when using the same AFM area in both cases.

Assuming identical AFM areas in cases (ii) and (iii), the interaction strength at resonance frequency $\omega$ between cavity modes of momentum $\vq$ and the uniform magnon mode for both cases is approximately 
\begin{align}
    \tilde{g} \approx {}& \frac{e m\Area_\mathrm{AFM}(u_0 + v_0)}{8\pi \vF } \sqrt{\frac{\omega}{\eps_0 S N \Vol}}\nonumber\\*
    \approx {}& \frac{em}{4\pi \vF}\sqrt{\frac{\omega_E \Area_\mathrm{uc}}{2\eps_0 S L_z}} \sqrt{\frac{\Area_\mathrm{AFM}}{\Area_\mathrm{cav}}},
\end{align}
where we have used $\Vol = L_z \Area_\mathrm{cav}$ and $\Area_\mathrm{AFM}/N = \Area_\mathrm{uc}$. In the limit $\Area_\mathrm{AFM} \to \Area_\mathrm{cav}$, the above is in perfect agreement with the result in the main text. Moreover, since $\Area_\mathrm{AFM} \propto N$, the above coupling scales as $\sqrt{N}$, as is often expected for magnon-photon coupling. This scaling is not present when the subsystem areas are all identical, in which case there is no way to increase the number of spins except by altering material parameters.

\begin{acknowledgments}
\paragraph{Acknowledgments.}
We acknowledge funding via the ``Outstanding Academic Fellows'' programme at NTNU and the Research Council of Norway Grant Nos. 302315, 262633 and 331821.
\end{acknowledgments}


%

\end{document}